\documentclass{elsart}

 \usepackage{graphicx}

\usepackage{amssymb}

\begin{document}

\begin{frontmatter}



\title{Low energy neutron propagation in MCNPX and GEANT4}

 \author[a]{R. Lemrani}
 \author[b]{M. Robinson}
 \author[b]{V. A. Kudryavtsev}
 \author[c]{M. De Jesus}
 \author[a]{G. Gerbier}
 \author[b]{ N. J. C. Spooner}
 \address[a]{DAPNIA-SPP, CEA-Saclay, Gif-Sur-Yvette, France}
 \address[b]{Department of Physics and Astronomy, University of Sheffield, S3 7RH, UK}
 \address[c]{Institut de Physique Nucl\'eaire de Lyon, IN2P3-CNRS, Universit\'e Claude
Bernard Lyon-I, 4 rue Enrico Fermi, 69622 Villeurbanne Cedex, France}


\begin{abstract}

Simulations of neutron background from rock for underground experiments are presented. 
Neutron propagation through two types of rock, lead and hydrocarbon material 
is discussed. The results show a reasonably good agreement between 
GEANT4, MCNPX and GEANT3 in transporting low-energy neutrons.  
\end{abstract}

\end{frontmatter}

\section{Introduction }
Neutrons are known to be an important source of background for high sensitivity experiments searching for rare 
events in underground laboratories. These experiments include the direct dark matter searches, double-beta decay 
experiments, solar neutrino measurements etc. Significant progress has been achieved in the past few years in the 
development of techniques for rare event searches at different energies. This imposes more stringent restrictions on the tolerated background rates.

Dark matter direct detection experiments have developed over the last decade strong background suppression and
rejection techniques. These experiments are installed in deep underground laboratories (Gran Sasso, Modane, Canfranc,
Boulby, Soudan etc.) reducing cosmic ray related backgrounds by many orders of magnitude. Lead and copper shielding
is used against gamma-rays. Internal contamination is kept low by the selection of low radioactive materials.
Materials implemented in the vicinity of the detectors are purified for instance by chemical or mechanical cleaning
to achieve low radioactivity. Cryogenic detectors have shown a powerful event-by-event discrimination of electron and
nuclear recoils \cite{aker,sang,angl}. Liquid noble gas detectors (see, for example, Refs.
\cite{alne,apri}) and
other techniques are
progressing towards effective gamma rejection. With the sensitivities thus achieved neutrons have become the limiting background. 

Unlike other backgrounds (photons, electrons, alphas) neutrons may induce low-energy single nuclear recoils in the
detectors indistinguishable from the expected WIMP signal. In underground laboratories neutrons originate
overwhelmingly from the presence of uranium and thorium contaminations in the surrounding rock. Uranium and thorium
produce neutrons via spontaneous fission (mainly $^{238}$U) or indirectly via interactions of alphas in their
decay
chains with the light nuclei of the rock. Neutrons induced by cosmic-ray muon interactions with rock and shielding 
have much lower rate but higher probability to reach the detectors. Active veto systems will help to reject these events if necessary.

To probe the region of parameter space favoured by SUSY models a sensitivity down to at least  10$^{-10}$ pb to the
WIMP-nucleon cross-section is required. This is translated to only a few events per year per tonne of the target
material in the energy range of interest. To reach such a sensitivity, the neutron background from rock, detector
components and cosmic-ray muons should be significantly suppressed. For example, the neutron flux from rock should be
reduced by at least 6 orders of magnitude. This can be achieved by installing passive shielding made of hydrocarbon
material or water around the target. To design the shielding, its thickness, configuration and composition, Monte
Carlo simulations are required. As we are talking here about several orders of magnitude reduction in the neutron
flux passing through the shielding, we have to be sure that the Monte Carlo codes are accurate enough for such a job.
It is therefore crucial to assess the reliability of the simulations used to design the shielding of future
experiments. This paper compares for the first time propagation of neutrons through large thickness of rock and
shielding using the Monte Carlo codes MCNPX \cite{brie,mcnp} and GEANT4 \cite{agos}. Some simulations were
also done with
GEANT3
\cite{gean}. The calculation of the neutron production spectrum in different rocks is described in Section 2. In
Section 3
the propagation of neutrons to the rock/laboratory boundary and through various shielding materials is presented. In
Section 4 we compare the predictions of different Monte Carlo codes.

\section{Neutron production in rock }
The present study has been performed for two types of rock: NaCl, which is the rock around the Boulby Underground
Laboratory hosting the ZEPLIN \cite{alne} and DRIFT \cite{alne2} experiments, and the Modane rock, the Modane
Underground
Laboratory
(LSM) being the site for the EDELWEISS \cite{sang} and NEMO-3 \cite{sima} experiments. NaCl is also the rock
around
the
WIPP –
site
suggested for several future experiments. There is a plan to construct another large laboratory for underground 
science in the Frejus tunnel which hosts the Modane laboratory. Thus the present study is relevant to several 
existing and future experiments and also provides important general outcomes about the code accuracy and the neutron suppression factors in the shielding.

The neutron production rates and spectra in NaCl were calculated in \cite{cars} with the SOURCES code
\cite{wils}.
The
code
determines the energy spectrum of neutrons produced in ($\alpha$,n) reactions, spontaneous fission and delayed 
neutron emission due to radioactive isotopes present in the rock. The code uses Watt spectrum parameters, evaluated 
half-life and spontaneous fission branching as input for the evaluation of the spontaneous fission contribution. 
The ($\alpha$,n) spectra assume an isotropic angular distribution in the center-of-mass system. The alpha stopping 
power in various media, evaluated or measured ($\alpha$,n) cross-section tables and branching ratios for transitions 
to different excited states are used to calculate the thick target neutron yields and neutron spectra. 

The code SOURCES was modified \cite{cars} to allow calculation of neutron yield from high-energy (more than 6.5
MeV)
alphas.
Some cross-sections were updated and more cross-sections added to the code library. The resulting spectrum assuming
60 ppb U and 300 ppb Th concentrations in NaCl is shown in Figure 1  (see also \cite{cars}).

The modified SOURCES code has also been used in the present study for the evaluation of the neutron production rate
in the Modane Underground Laboratory (LSM). The LSM rock consists of glossy schist (2.65 g/cm$^3$ density), the
element concentrations (O - 50\% , Ca - 31\%, C - 6\%, Si - 7\%, Mg - 1\%, Al - 2\%, H - 1\%, Fe - 2\%) have been
determined by the spectroscopy analysis \cite{chaz}. New Ca and Mg cross-sections computed using the code EMPIRE
\cite{empi}
have
been included in the SOURCES cross-section library. The concentrations of 0.84 ppm $^{238}$U and 2.45 ppm $^{232}$Th
have been assumed in the calculation of the neutron production rate. A production rate of 1.3 neutrons/g/year has
been obtained with a mean energy of 2.2 MeV (see Figure 1). Despite higher radioactivity levels for Modane rock
compared to what was assumed for NaCl, the neutron production rates are very similar (1.5 neutrons/g/year in NaCl
\cite{cars}). This is due to the higher ($\alpha$,n) contribution per unit U/Th concentration in NaCl because of
low
energy thresholds for $^{23}$Na (3.5 MeV) and $^{37}$Cl (4.1 MeV), the spontaneous fission yield per unit U 
concentration being the same for both rocks. This results in $98\%$ contribution of ($\alpha$,n) reactions 
in NaCl and only $77\%$ contribution of ($\alpha$,n) reactions in the Modane rock.

\section{Simulations of the neutron propagation }
The neutron propagation calculations have been carried out using the MCNPX-2.5 \cite{brie,mcnp} and
GEANT4.7.0.p01 \cite{agos}
simulation codes. High precision model for low-energy neutron tracking has been chosen in the GEANT4 simulations
using neutron cross-section library G4NDL3.7. In
MCNPX the libraries NRG-2003, la150n and JENDL \cite{mcnp} were used for lead, CH$_2$ and NaCl respectively. Some
test
simulations have also been run with GEANT3 \cite{gean}. Starting from the neutron production spectra provided by
the
SOURCES
code, the neutrons have first been propagated to the rock boundary and then through different configurations of
shielding. For the benchmark a simple geometry has been used, where the boundaries between different media are planar
(see Figure 2). 

Neutrons have been generated in a volume of rock of 1 $\times$  1 m$^2$ section and 3 m depth as there were
essentially no
neutrons crossing more than 3 metres of rock. The volume for neutron propagation has been taken much larger with a
cross-section of 10 $\times$  10 m$^2$ and 3 m depth. This has allowed collection of all neutrons which could reach
the
rock/cavern boundary. Two configurations of shielding have been studied. In the first one, neutrons have been
propagated through different thicknesses (5 g/cm$^2$ to 50 g/cm$^2$) of hydrocarbon shielding CH$_2$. In the second
one, neutrons have first been propagated through a slab of lead 30 cm thick before further propagation through
CH$_2$.
Lead is commonly used in underground laboratories as gamma shielding so it was worth studying the neutron propagation
in lead and its effect on the neutron flux suppression. Note that the chosen geometry is the same as used in Ref.
\cite{cars} allowing direct comparison of the results. In Ref. \cite{cars} the simulations were done for NaCl
with
GEANT4.5.2.

All neutrons have been counted on each surface and the fluxes have been normalised to the area of 1 m$^2$. In this 
way we can say that we counted all neutrons generated in a volume with a surface projection cross-section of 1 m$^2$, 
that reached the surface of a medium. Other cells with the same projection cross-section will give the same neutron flux.

\section{Results}
Figure 3 shows the simulated neutron spectra at the surface of NaCl. The total statistics for each curve on
this and subsequent graphs is at least $10^5$ neutrons leading to more than 1000 neutrons per energy bin 
at all energies between 10 keV and 2 MeV.
GEANT4, MCNPX and GEANT3 simulations differ at
most by $20\%$ in a narrow region around 1 MeV. The integrated neutron fluxes above 1 MeV agree within $10\%$. Note
that the GEANT4 results obtained with the old cross-section library and published in Refs. \cite{cars,cars2} differ from the
present GEANT4 simulations above 1 MeV. The difference is due to an error found in the inelastic cross-section on
chlorine \cite{araj} and corrected in the present simulations. The smaller neutron flux above 1 MeV obtained with the
corrected cross-section, results in smaller fluxes by about a factor of 2-3 after 20-40 g/cm$^2$ of CH$_2$ shielding. 
Agreement between GEANT4 and MCNPX for the Modane rock (Figure 4) is again fairly good, the maximal difference being about $15\%$ around 2 MeV. 

The neutron flux in the Modane rock is smaller than in NaCl despite higher U/Th concentrations and similar neutron
production rates (discussed above). This is due to the presence of a small amount of hydrogen (in water) in the
Modane rock and its absence in NaCl. Hydrogen serves as a good neutron moderator and the flux is very sensitive to
the hydrogen abundance. Hydrogen reduces the neutron flux above 100 keV (1 MeV) at Modane by a factor 4.7
(1.8)  (see also Ref. \cite{wula} for a discussion of this effect in concrete). 

The neutron spectra originated in NaCl and propagated through different thicknesses of CH$_2$ shielding are shown in
Figure 5. The agreement is satisfactory for the purpose of designing shielding. After 50 g/cm$^2$ of CH$_2$, MCNPX
gives a differential flux $50\%$ higher than the GEANT4 flux, translating into an additional 1-2 cm thick CH$_2$
layer. Figure 6 shows the corresponding integrated fluxes above 100 keV and 1 MeV. The results for GEANT4 simulations
are slightly different from published in Ref. \cite{cars} due to the corrected inelastic cross-section on chlorine
used
in the present study. The results of neutron propagation through 30 cm of Pb and additional CH$_2$ layers are shown 
in Figures 7 and 8. Again a reasonable agreement between MCNPX and GEANT4 is observed. 

To calculate accurately the real flux in a laboratory one would have to take into account the exact
geometry of the cavern since the back-scattering of neutrons off the cavern walls increases the neutron flux
significantly (see the discussion in Refs. \cite{cars,smit}). Figure 9 shows the direct neutron flux coming from the walls of NaCl rock to the cavern with the size of 30 $\times$ 6.5 $\times$ 4.5 m$^{3}$ (same as in Ref. \cite{cars}) and the
total flux taking 
into account the back-scattering of neutrons. Both GEANT4 and MCNPX predict a flux enhancement due to back-scattering of 50$\%$ above 1 MeV.
Similarly in the case of the Modane underground laboratory (30 $\times$ 11 $\times$ 10 m$^{3}$) a flux enhancement of
32$\%$ above 1 MeV is obtained with GEANT4 and 25$\%$ with MCNPX.

\section{Conclusions}
We have shown the first comparison of MCNPX, GEANT4 and GEANT3 simulations of neutron propagation through different materials
 relevant to underground experiments. A reasonably good agreement is observed for two types of rock, lead and hydrocarbon 
material. The maximal observed difference in differential and integrated neutron fluxes is of the order of $50\%$ after 50 g/cm$^2$ 
of CH$_2$ with MCNPX giving a flux higher than GEANT4. Keeping in mind that 
the flux attenuation in 50 g/cm$^2$ of CH$_2$ is about 6 orders of magnitude,
 such a difference does not give any reason to question the reliability of MCNPX and GEANT4 codes. 
This result provides a solid ground for the design of neutron shielding for experiments aiming at a background free environment. 
From Figures 5-8 we conclude that a neutron shielding equivalent to 55 g/cm$^2$ should suppress the neutron flux from environment radioactive origin by more than 6 orders of magnitude. This is roughly within the requirements  for a tonne-scale dark matter experiment aiming at sensitivity of $10^{-10}$~pb to WIMP-nucleon spin-independent interactions \cite{cars}. Exact composition, thickness and configuration of the shielding will be determined on the basis of specific features of a particular set up. 
The results can also be used as a benchmark for those who would want to check their own simulations.

\section{Acknowledgements}
This work has been partially supported by the ILIAS integrating activity (Contract No. RII3-CT-2004-506222) as part of the EU FP6 programme in Astroparticle Physics.



\newpage

\begin{figure}
\begin{center}
\includegraphics[width=3.5in]{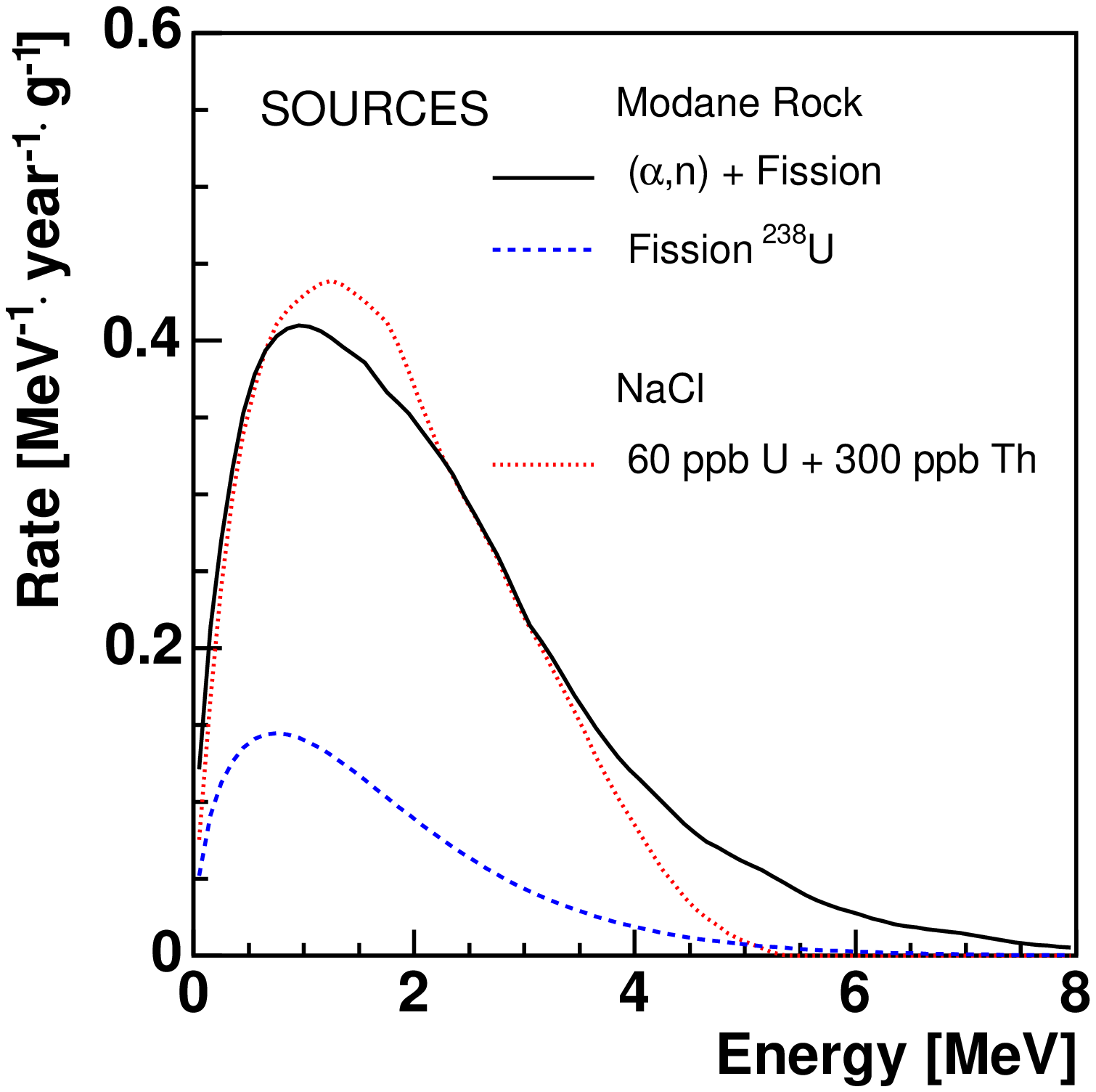}
\caption{Neutron energy spectrum from U and Th traces in the Modane rock simulated with SOURCES (full line)
and fission contribution  (dashed line). 
The spectrum for 60 ppb U + 300 ppb Th in NaCl is also shown (dotted line).  }
\end{center}
\end{figure}
\begin{figure}
\begin{center}
\includegraphics[width=3.5in]{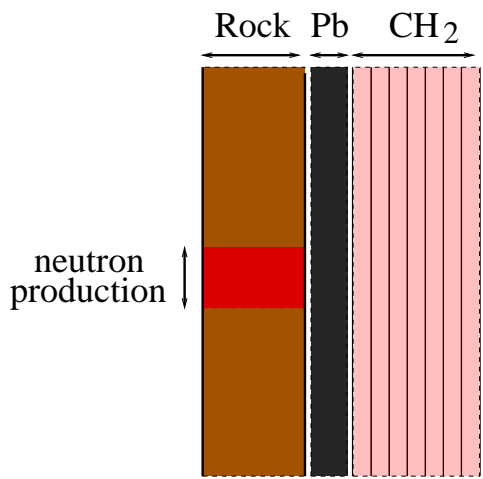}
\caption{ Sketch of the geometry used as a benchmark for comparing the simulations. }
\end{center}
\end{figure}
\begin{figure}
\begin{center}
\includegraphics[width=3.5in]{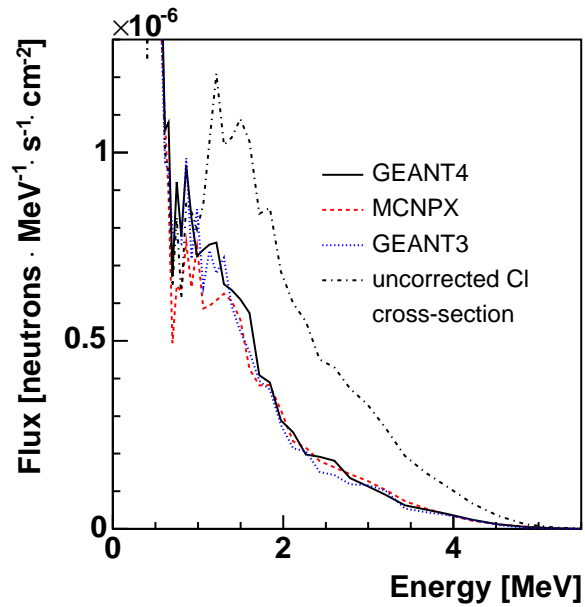}
\caption{ Neutron energy spectra from U and Th traces in NaCl at rock boundary simulated 
with GEANT4 (solid line), MCNPX (dashed line) and GEANT3 (dotted line). 
The GEANT4 result with uncorrected inelastic cross-section is also shown (dashed-dotted line).}
\end{center}
\end{figure}
\begin{figure}
\begin{center}
\includegraphics[width=3.5in]{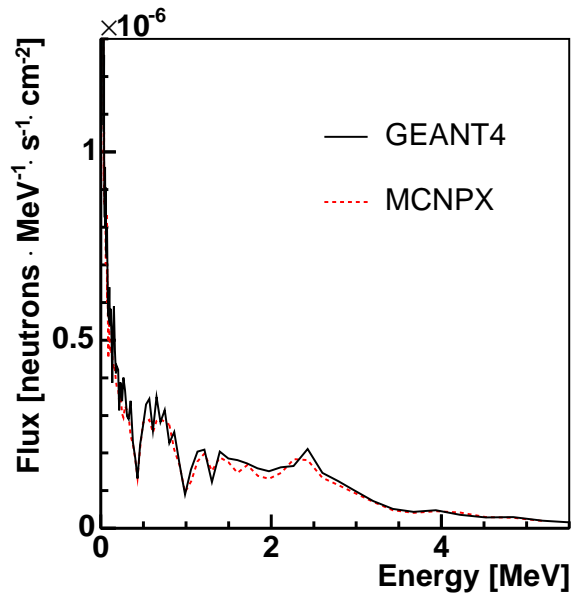}
\caption{Neutron energy spectra from rock activity at Modane rock boundary simulated with GEANT4 (solid line) and MCNPX (dashed line).}
\end{center}
\end{figure}
\begin{figure}
\begin{center}
\includegraphics[width=3.5in]{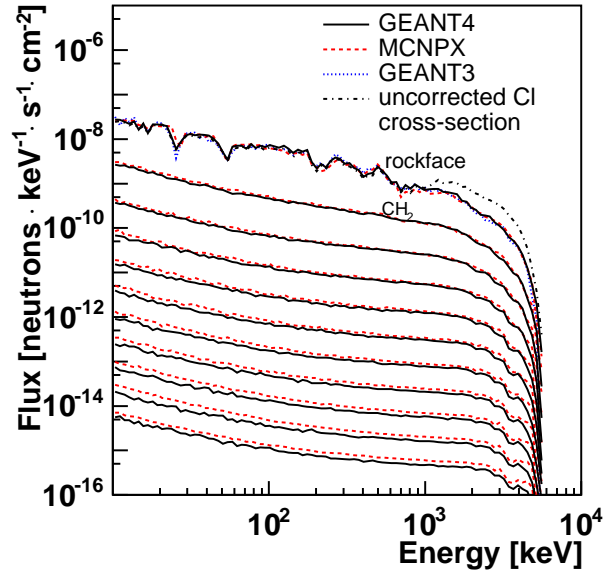}
\caption{ Neutron energy spectra from NaCl after hydrocarbon shielding simulated with GEANT4 (solid lines) and MCNPX
(dashed lines).The GEANT4 result with uncorrected inelastic cross-section is also shown (dashed-dotted line).}
\end{center}
\end{figure}
\begin{figure}
\begin{center}
\includegraphics[width=3.5in]{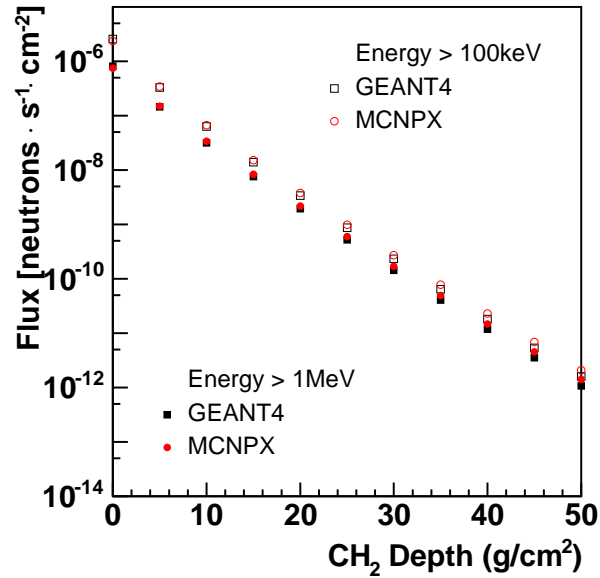}
\caption{Neutron flux from NaCl above 100 keV (open circles and squares) and above 1 MeV (full circles and squares)
as a function of CH$_2$ thickness simulated with MCNPX (circles) and GEANT4 (squares).}
\end{center}
\end{figure}
\begin{figure}
\begin{center}
\includegraphics[width=3.5in]{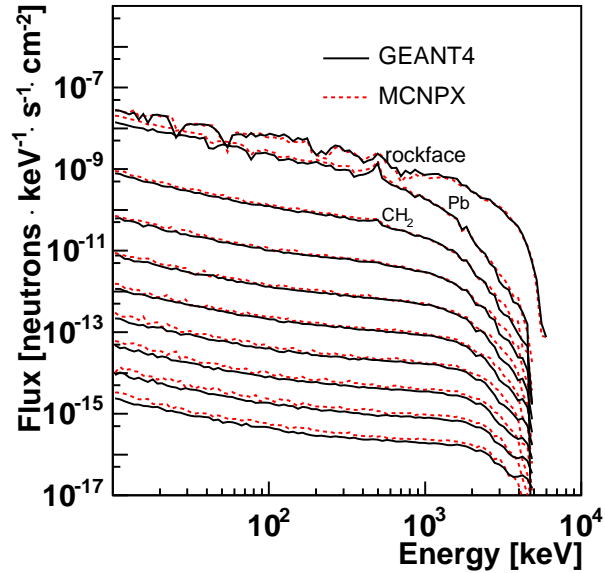}
\caption{ Neutron energy spectra from NaCl after lead and hydrocarbon shielding simulated with GEANT4 (solid lines) and MCNPX (dashed lines).}
\end{center}
\end{figure}
\begin{figure}
\begin{center}
\includegraphics[width=3.5in]{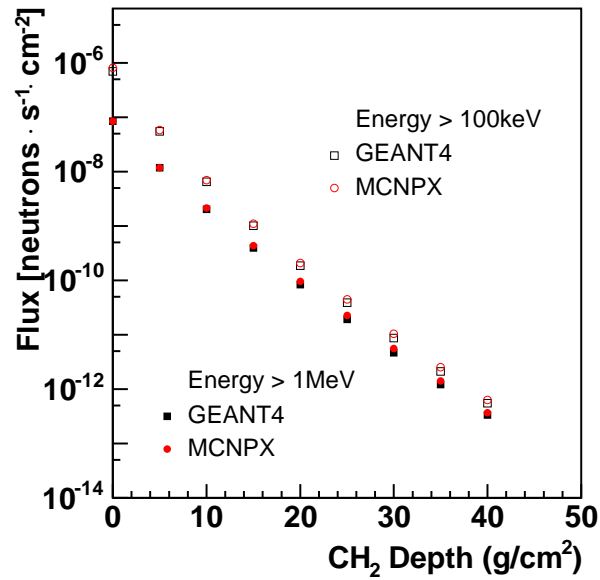}
\caption{ Total neutron flux from NaCl above 100 keV (open circles and squares) and above 1 MeV (full circles and
squares) as a function of CH$_2$ thickness after 30 cm of lead calculated with MCNPX (circles) and GEANT4 (squares).}
\end{center}
\end{figure}
\begin{figure}
\begin{center}
\includegraphics[width=3.5in]{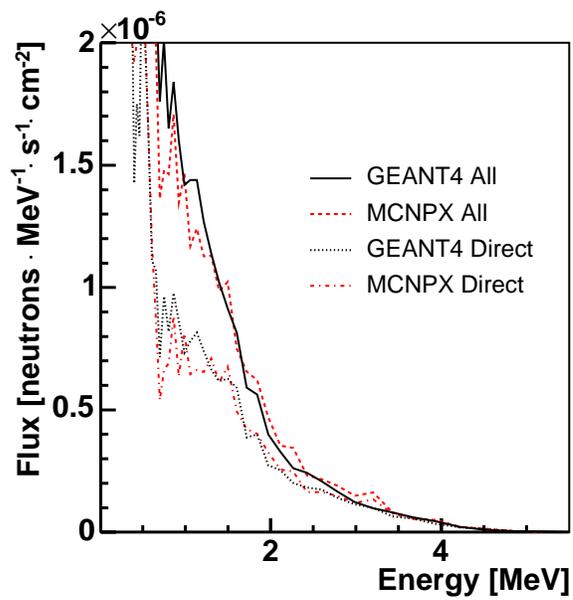}
\caption{ Neutron energy spectra at the rock/cavern boundary from U and Th traces in NaCl taking into account
back-scattering effect with GEANT4 (solid line) and MCNPX (dashed line). Also shown the direct flux (without 
back-scattering) with GEANT4 (dotted line) and MCNPX (dashed-dotted line).} 
\end{center}
\end{figure}

\end{document}